\documentstyle[12pt,psfig,paspconf]{article}
\begin{document}

{\large\bf{Chromospheric activity of
ROSAT discovered weak-lined T Tauri stars}}

{\it{ D. Montes$^{1,2}$, L.W. Ramsey$^{1}$
}}

$^1$ {The Pennsylvania State University,
Department of Astronomy and Astrophysics,
525 Davey Laboratory, University Park, PA 16802, USA}\\
$^2$ {Departamento de Astrof\'{\i}sica,
Facultad de F\'{\i}sicas,
Universidad Complutense de Madrid, E-28040 Madrid, Spain}

\vspace*{0.6cm}

To be published  in ASP Conf. Ser., Solar and Stellar Activity: 
Similarities and Differences
(meeting dedicated to Brendan Byrne, Armagh 2-4th September 1998) 
C.J. Butler and J.G. Doyle, eds

\vspace*{2.5cm}

\baselineskip=0.6truecm

\large

\hrule
\vspace{0.2cm}
\begin{center}
{\huge\bf Abstract}
\end{center}

We have started a
high resolution optical observation
program dedicated to the study
of chromospheric activity
in weak-lined T Tauri stars (WTTS) recently
discovered by the ROSAT All-Sky Survey (RASS).
It is our purpose to
quantify the phenomenology of the chromospheric activity of each star
determining stellar surface fluxes in the more important
chromospheric activity indicators
(Ca~{\sc ii} H \& K, H$\beta$, H$\alpha$, Ca~{\sc ii} IRT)
as well as obtain the Li~{\sc i} abundance, a
better determination of the stellar parameters, spectral type,
and possible binarity.
With this information we can
study in detail
the flux-flux and rotation-activity relations
for this kind of objects and compare it with the
corresponding relations in the  well studied
RS CVn systems.

A large number of WTTS have been discovered by the RASS in
and around different star formation clouds.
Whether these stars are really WTTS, or post-TTS, or
even young main sequence stars is a matter of ongoing debate.
However, we have centered our study only on objects
for which very recent studies, of Li~{\sc i} abundance
(greater than Pleiads of the same spectral type)
or radio properties,
clearly confirmed their pre-main sequence (PMS) nature.

In this contribution we present preliminary results
of our January 1998 high resolution echelle spectroscopic observations
at the 2.1m telescope of the  McDonald Observatory.
We  have analysed, using  the spectral subtraction technique,
the H$\alpha$ and Ca~{\sc ii} IRT lines of six WTTS
(RXJ0312.8-0414NW, SE; RXJ0333.1+1036; RXJ0348.5+0832; RXJ0512.0+1020;
RXJ0444.9+2717)
located in and around the Taurus-Auriga molecular clouds.
A broad and variable double-picked H$\alpha$ emission is observed in
RXJ0444.9+2717.
Emission above the continuum in H$\alpha$ and Ca~{\sc ii} IRT lines
is detected in RXJ0333.1+1036
and a filling-in of these lines is present in the rest of the stars.
Our spectral type and Li~{\sc i} EW deterninations
confirm the PMS nature of these objects.

\vspace{0.4cm}
\hrule

\newpage
\hrule
\vspace{0.2cm}
\begin{center}
{\huge\bf Introduction}
\end{center}

Weak-lined T Tauri stars (WTTS) are low mass pre-main sequence stars (PMS)
with H$\alpha$ equivalent widths $\leq$ 10 {\AA} in which no
signs of accretion are observed.
The emission spectrum of these stars is not affected
by the complications of star-disk interaction which
often masks the underlying absorption lines as well as extincts the stellar
light in classical T Tauri stars (CTTS).
The WTTS are thus ideal targets to study the behavior of surface
activity in the PMS stage of the stellar evolution.
While there are a large number of studies in UV, X-ray and radio
wavelengths, little research has been directed towards the study of the
chromospheric activity using optical observations.
Those which have been done
are based on low resolution spectroscopic observations.
Only some recent higher resolution studies centered in bona-fide WTTS
in Taurus are available
(see Feigelson et al. 1994; Welty 1995; Welty \& Ramsey 1995, 1998;
Poncet et al. 1998; Montes \& Miranda 1999).
In order to improve the knowledge of the WTTS chromospheres
high resolution optical observations are needed.
The WTTS discovered very recently by the ROSAT All-Sky Survey (RASS)
are good targets to accomplish these objectives.
A large number of them have been found far away from the
star formation clouds (Neuh\"auser et al. 1995; Alcal\'{a} et al. 1995, 1996;
Wichmann et al. 1996; Magazz\`{u} et al. 1997; Krautter et al. 1997).
Whether these stars are really WTTS, or post TTS, or
even young main sequence stars is a matter of ongoing debate
(Feigelson 1996, Brice\~{n}o et al. 1997, Favata et al. 1997).
However, we will study
only those in the Taurus Auriga Molecular Cloud
for which very recent studies
clearly confirmed their PMS nature.

In this contribution we present preliminary results
of our high resolution echelle spectroscopic observations
of RX~J0312.8-0414NW, SE; RX~J0333.1+1036; RX J0348.5+0832; RX J0512.0+1020;
and RX J0444.9+2717.

\vspace{0.4cm}
\hrule

\newpage
\hrule
\vspace{0.2cm}
\begin{center}
{\huge\bf Observations}
\end{center}

The spectroscopic observations were obtained
during a 10 night run 12-21 January 1998 using the 2.1m telescope at
McDonald Observatory and the Sandiford Cassegrain Echelle Spectrograph
(McCarthy et al. 1993).
This instrument is a prism crossed-dispersed echelle mounted at the
Cassegrain focus and it is used with a 1200X400 Reticon CCD.
The spectrograph setup was chosen to cover the
H$\alpha$ (6563~\AA) and Ca~{\sc ii} IRT (8498, 8542, 8662~\AA) lines.
The wavelength coverage is about 6400-8800\AA$\ $ and the
reciprocal dispersion ranges from 0.06 to 0.08 ~\AA/pixel.
The spectral resolution, determined by the FWHM of the arc
comparison lines, ranges from 0.13 to 0.20 \AA$\ $
(resolving power R=$\lambda$/$\Delta\lambda$ of 50000 to 31000)
in the H$\alpha$ line region.
In one of the nights we changed the spectrograph setup to include the
He{\sc i} D$_{3}$ (5876~\AA)
and Na~{\sc i} D$_{1}$ and D$_{2}$ (5896, 5890~\AA)
with a wavelength coverage of 5600-7000~\AA.
The spectra have been extracted using the standard
reduction procedures in the IRAF
 package (bias subtraction,
flat-field division, and optimal extraction of the spectra).
The wavelength calibration was obtained by taking
spectra of a Th-Ar lamp.
Finally, the spectra have been normalized by
a low-order polynomial fit to the observed continuum.
The chromospheric contribution in
these features is determined  using
the spectral subtraction technique
(Huenemoerder \& Ramsey 1987; Montes et al. 1995; 1997).
The synthesized spectrum was constructed using the program STARMOD
developed at Penn State (Barden 1985).

\vspace{0.4cm}
\hrule

\newpage
\hrule
\vspace{0.2cm}
\begin{center}
{\huge\bf Results}
\end{center}

We  have analysed
the H$\alpha$ and Ca~{\sc ii} IRT lines of six WTTS
(RX~J0312.8-0414NW, SE; RX~J0333.1+1036; RX~J0348.5+0832; RX~J0512.0+1020;
RX J0444.9+2717)
located in and around the Taurus-Auriga molecular clouds.
These targets were selected from two sources:

(1) From the ROSAT detected late-type stars
south of the Taurus
(Neuh\"auser et al. 1995; Magazz\`{u} et al. 1997 (hereafter M97))
we selected the stars that the spectroscopic studies
of  M97 and Neuh\"auser et al. (1997, hereafter N97)
clearly identified as WTTS from their greater Li abundance
than Pleiades of the same spectral type.
Some of them have been classified by these authors as
single- and double-lined spectroscopic binaries (SB1, and SB2)
and others are visual binaries (Sterzik et al. 1997, hereafter S97).

(2) From the list of
Wichmann et al. (1996) (hereafter W96) of new WTTS stars in Taurus,
we selected the stars in which radio emission
was detected by Carkner et al. (1997, hereafter C97)
supporting their identification
as genuine WTTS rather than ZANS.

\vspace{0.2cm}

Representative spectra of these stars are plotted in
Fig.~1 (H$\alpha$), Fig.~2 (Li~{\sc i} 6708 \AA),
and Fig.~3 (Ca~{\sc ii} IRT).
A K1V reference star is also plotted for comparison.
The observed and subtracted spectra for the case of RX~J0444.9+2717 
are plotted in Fig.~4. 

\vspace{0.5cm}
\hrule

\newpage
\hrule
\vspace{0.2cm}

\vspace{0.6cm}

{\bf\underline{RX~J0312.8-0414NW, SE}}

\vspace{0.2cm}

RX~J0312.8-0414 is a visual binary with components NW and SE separated by 14''
(M97, S97). The NW component is a G0V with
{\it v}sin{\it i} = 33~km~s$^{-1}$ and is a SB2.
The SE component is a G8V with {\it v}sin{\it i} = 11~km~s$^{-1}$.
Both components exhibit H$\alpha$ absorption  with a                                      
EW(H$\alpha$) of 3.5 and of of 2.5 \AA\ respectively (M97; N97).

Our spectra exhibit a strong Li~{\sc i} 6708 \AA\ line confirming
the PMS nature of these objects.
However, the level of chromospheric activity is very low, only
a small filling-in of H$\alpha$ and Ca~{\sc ii} IRT lines is detected,
in agreement with the earlier spectral type of both stars.

\vspace{0.6cm}

{\bf\underline{RX~J0333.1+1036}}

\vspace{0.2cm}

This star is classified as a confirmed PMS star by M97 and N97
on the basis of
its Li~{\sc i} abundance, however, C97 detect no radio emission.
M97 give a spectral type of K3 and observed the H$\alpha$ line in emission
with a EW of -0.8 \AA. N97 measured a {\it v}sin{\it i} of 20~km~s$^{-1}$

Emission above the continuum in H$\alpha$ and Ca~{\sc ii} IRT lines
is detected in our five spectra from January 14 to January 20 1998
with small variations from night to night.

\vspace{0.6cm}

{\bf\underline{RX~J0348.5+0832}}

\vspace{0.2cm}

This PMS is a rapidly-rotating star
({\it v}sin{\it i} = 127 km~s$^{-1}$, N97)
of spectral type G7 and with a small emission in the H$\alpha$ line
(EW = -0.1 \AA, M97).

In our five spectra (from 01/12/98 to 01/18/98) we observe the H$\alpha$
line always in absorption, but filled in.
The Ca~{\sc ii} IRT lines are also filled in by chromospheric emission.

\vspace{0.6cm}

{\bf\underline{RX~J0512.0+1020}}

M97 give a spectral type K2 for this star and observed a
small emission in the H$\alpha$ line
(EW = -0.1 \AA).
N97 measured a rotational velocity of 57~km~s$^{-1}$.

In our three spectra (from 01/14/98 to 01/17/98) we observe a
variable
filling-in of the H$\alpha$ and Ca~{\sc ii} IRT lines.
The H$\alpha$ line shows emission in the blue wing in one of the spectra.

\vspace{0.5cm}
\hrule

\newpage
\hrule
\vspace{0.2cm}

\vspace{0.6cm}

{\bf\underline{RX~J0444.9+2717}}

\vspace{0.2cm}

This is a K1 star with H$\alpha$ emission above the continuum
(EW = -2.1 \AA) and classified as a PMS star by W96
on the basis of its Li~{\sc i} abundance.
The detection of radio emission by C97 confirm its PMS nature.
Kohler \& Leinert (1998) found a IR companion with a separation of 1.754'' and
a brignness ratio at K of 0.102.

We have eight spectra of this star available (from 01/12/98 to 01/20/98).
The observed spectra are well matched using a K1V reference star
with a rotational broadening of 65~km~s$^{-1}$.
Some of the more intense photospheric lines exhibit a flat-bottomed core
(i.e. the core is noticeable filled in with respect to the reference profile)
as is observed in other rapidly-rotating and spotted stars.
A broad and variable double-picked H$\alpha$ emission above the continuum
is observed (see Fig.~4).
The H$\alpha$ EW in the observed spectra changes from -1.2~\AA\  to -2.6~\AA.
The  Ca~{\sc ii} IRT lines exhibit a strong filling-in.

\vspace{0.4cm}
\hrule


\clearpage

\begin{figure*}
{\psfig{figure=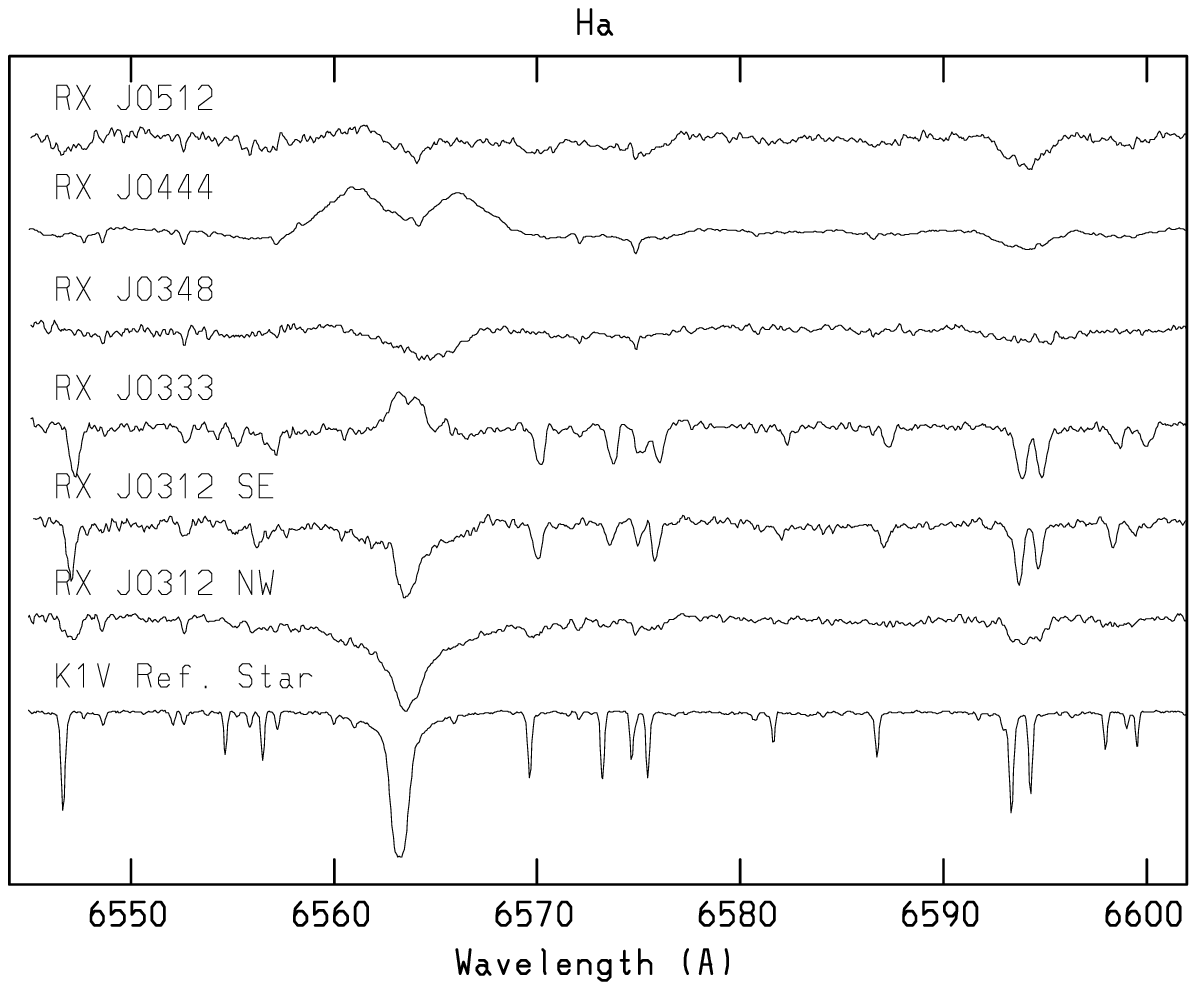,bbllx=128pt,bblly=253pt,bburx=480pt,bbury=542pt,height=17.5cm,width=18.6cm,clip=}}
\caption[ ]{Spectra in the region of the H$\alpha$ line}
\end{figure*}

\begin{figure*}
{\psfig{figure=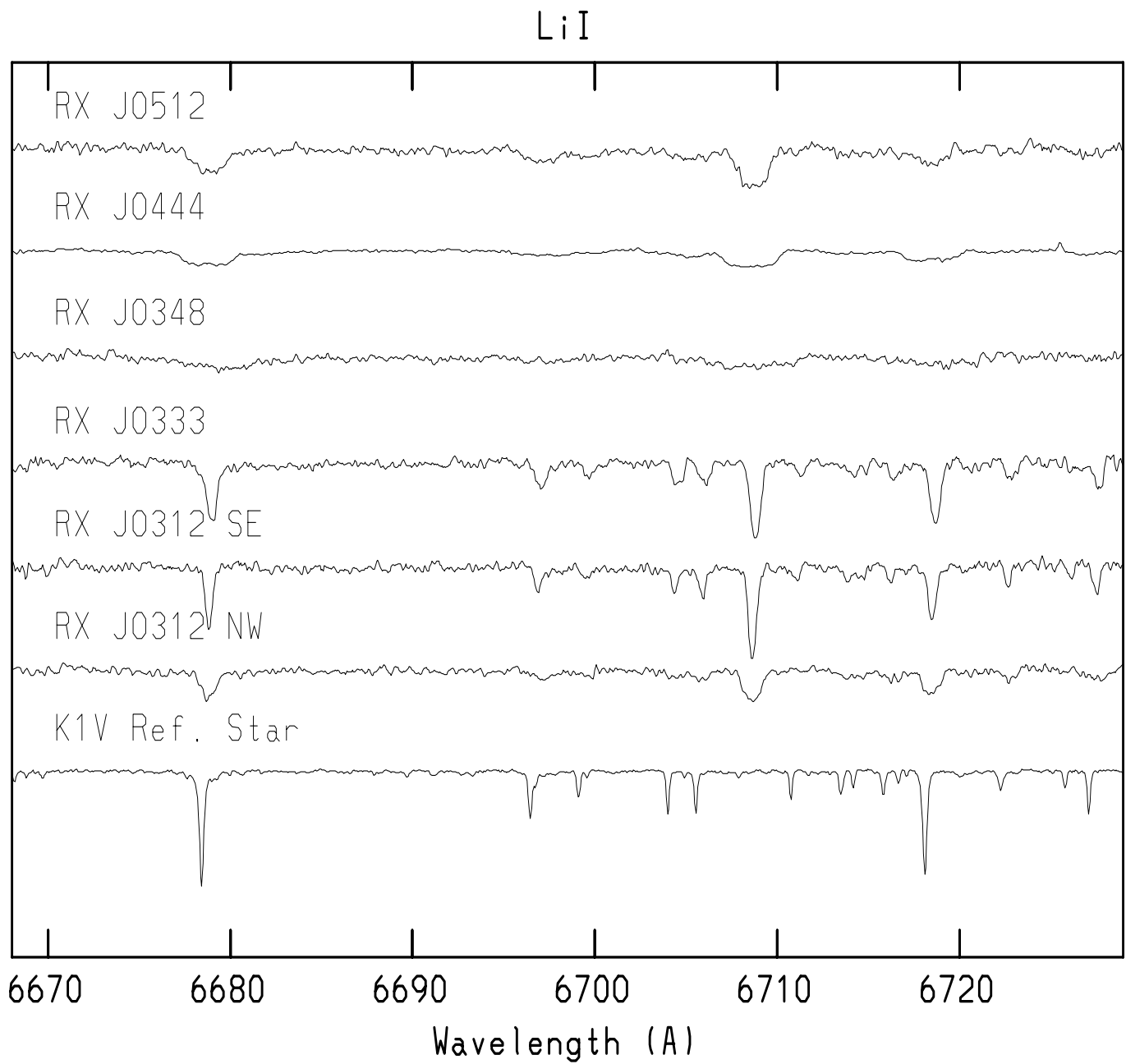,bbllx=128pt,bblly=253pt,bburx=480pt,bbury=542pt,height=17.5cm,width=18.6cm,clip=}}
\caption[ ]{Spectra in the region of the Li~{\sc i} $\lambda$ 6708~\AA\ line}
\end{figure*}

\begin{figure*}
{\psfig{figure=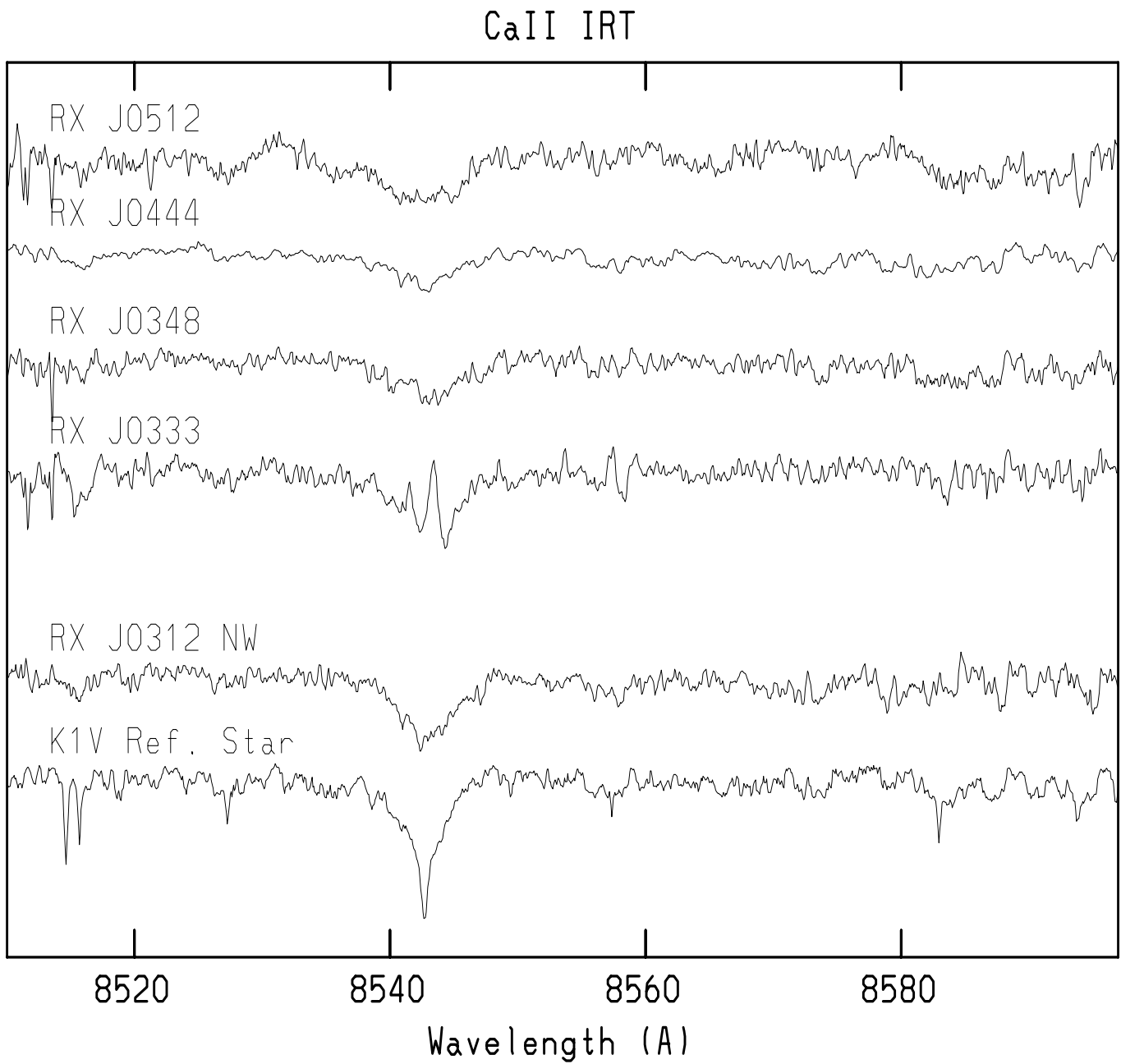,bbllx=128pt,bblly=253pt,bburx=480pt,bbury=542pt,height=17.5cm,width=18.6cm,clip=}}
\caption[ ]{Spectra in the region of the Ca~{\sc ii} IRT $\lambda$ 8542~\AA\
line}
\end{figure*}

\begin{figure*}
{\psfig{figure=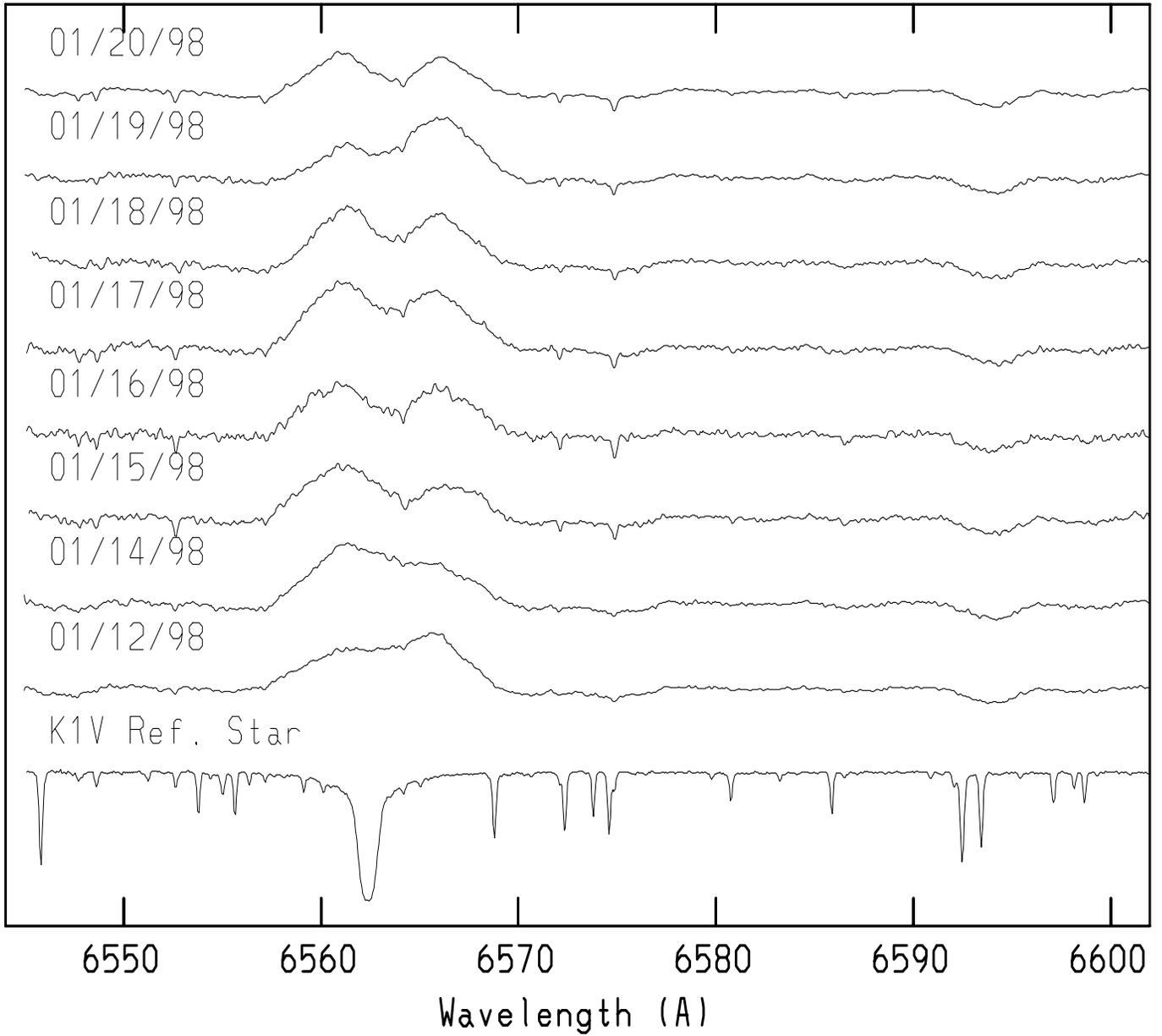,bbllx=128pt,bblly=253pt,bburx=480pt,bbury=542pt,height=17.5cm,width=18.6cm,clip=}}
\caption[ ]{Observed spectra of RX~J0444.9+2717 in the H$\alpha$ line region 
}
\end{figure*}

\begin{figure*}
{\psfig{figure=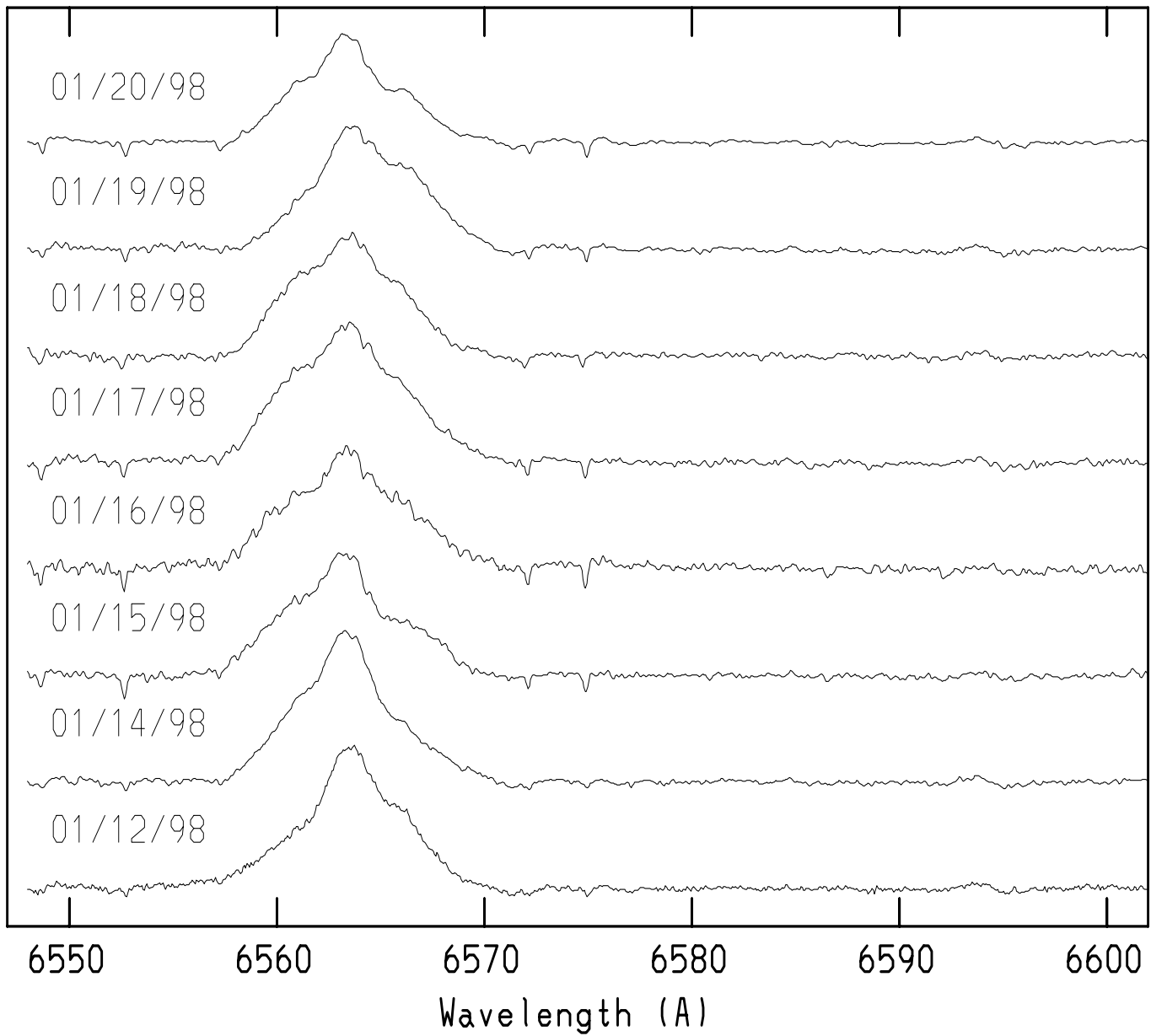,bbllx=128pt,bblly=253pt,bburx=480pt,bbury=542pt
,height=17.5cm,width=18.6cm,clip=}}
\caption[ ]{Subtracted spectra of RX~J0444.9+2717 in the H$\alpha$ line region
}
\end{figure*}


\end{document}